\documentclass[acmsmall]{acmart}
\usepackage{color,soul}
\usepackage{subcaption}
\usepackage{footnote}  
\AtBeginDocument{%
  \providecommand\BibTeX{{%
    \normalfont B\kern-0.5em{\scshape i\kern-0.25em b}\kern-0.8em\TeX}}}

\setcopyright{acmcopyright}
\copyrightyear{2020}
\acmYear{2020}

\begin{document}

\title{Explainable Active Learning (XAL): Toward AI Explanations as Interfaces for Machine Teachers}


\author{Bhavya Ghai}
\affiliation{%
  \institution{Stony Brook University}
  \city{Stony Brook}
  \state{NY}
  \country{USA}}
\email{bghai@cs.stonybrook.edu}

\author{Q. Vera Liao
}
\authornote{Corresponding author.}
\affiliation{%
  \institution{IBM Research AI}
  \city{Yorktown}
  \state{NY}
  \country{USA}}
\email{vera.liao@ibm.com}

\author{Yunfeng Zhang}
\affiliation{%
  \institution{IBM Research AI}
  \streetaddress{Yorktown}
  \city{Yorktown}
  \state{NY}
  \country{USA}}
\email{zhangyun@us.ibm.com}

\author{Rachel K. E. Bellamy}
\affiliation{%
  \institution{IBM Research AI}
  \city{Yorktown}
  \state{NY}
  \country{USA}}
\email{rachel@us.ibm.com}

\author{Klaus Mueller}
\affiliation{%
  \institution{Stony Brook University}
  \city{Stony Brook}
  \state{NY}
  \country{USA}}
\email{mueller@cs.stonybrook.edu}  

\begin{abstract}

The wide adoption of Machine Learning (ML) technologies has created a growing demand for people who can train ML models. Some advocated the term ``machine teacher'' to refer to the role of people who inject domain knowledge into ML models. This ``teaching'' perspective emphasizes supporting the productivity and mental wellbeing of machine teachers through efficient learning algorithms and thoughtful design of human-AI interfaces. One promising learning paradigm is Active Learning (AL), by which the model intelligently selects instances to query a machine teacher for labels, so that the labeling workload could be largely reduced. However, in current AL settings, the human-AI interface remains minimal and opaque. A dearth of empirical studies further hinders us from developing teacher-friendly interfaces for AL algorithms. In this work, we begin considering AI explanations as a core element of the human-AI interface for teaching machines. When a human student learns, it is a common pattern to present one's own reasoning and solicit feedback from the teacher. When a ML model learns and still makes mistakes, the teacher ought to be able to understand the reasoning underlying its mistakes. When the model matures, the teacher should be able to recognize its progress in order to trust and feel confident about their teaching outcome. Toward this vision, we propose a novel paradigm of explainable active learning (XAL), by introducing techniques from the surging field of explainable AI (XAI) into an AL setting. We conducted an empirical study comparing the model learning outcomes, feedback content and experience with XAL, to that of traditional AL and coactive learning (providing the model's prediction without explanation). Our study shows benefits of AI explanation as interfaces for machine teaching--supporting trust calibration and enabling rich forms of teaching feedback, and potential drawbacks--anchoring effect with the model judgment and additional cognitive workload. Our study also reveals important individual factors that mediate a machine teacher's reception to AI explanations, including task knowledge, AI experience and Need for Cognition. By reflecting on the results, we suggest future directions and design implications for XAL, and more broadly, machine teaching through AI explanations. 
\end{abstract}



\maketitle

\section{Introduction}
While Machine Learning technologies are increasingly used in a wide variety of domains ranging from critical systems to everyday consumer products, currently only a small group of people with formal training possess the skills to develop these technologies. Supervised ML, the most common type of ML technology, is typically trained with knowledge input in the form of labeled instances, often produced by subject matter experts (SMEs). Current ML development process presents at least two problems. First, the work to produce thousands of instance labels is tedious and time-consuming, and can impose high development costs. Second, the acquisition of human knowledge input is isolated from other parts of ML development, and often has to go through asynchronous iterations with data scientists as the mediator. For example, seeing suboptimal model performance, a data scientist has to spend extensive time obtaining additional labeled data from the SMEs, or gathering other feedback which helps in feature engineering or other steps in the ML development process~\cite{amershi2014power,brooks2015featureinsight}. 

The research community and technology industry are working toward making ML more accessible through the recent movement of ``democratizing data science''~\cite{chou2014democratizing}. Among other efforts, interactive machine learning (iML) is a research field at the intersection of HCI and ML. iML work has produced a variety of tools and design guidelines~\cite{amershi2014power} that enable SMEs or end users to interactively drive the model towards desired behaviors so that the need for data scientists to mediate can be relieved. More recently, a new field of ``machine teaching" was called for to make the process of developing ML models as intuitive as teaching a student, with its emphasis on supporting ``the teacher and the teacher's interaction with
data''~\cite{simard2017machine}.

The technical ML community has worked on improving the efficiency of labeling work, for which Active Learning (AL) came to become a vivid research area. AL could reduce the labeling workload by having the model select instances to query a human annotator for labels. However, the interfaces to query human input are minimal in current AL settings, and there is surprisingly little work that studied how people interact with AL algorithms. Algorithmic work of AL assumes the human annotator to be an oracle that provides error-free labels~\cite{settles2009active}, while in reality annotation errors are commonplace and can be systematically biased by a particular AL setting. Without understanding and accommodating these patterns, AL algorithms can break down in practice. Moreover, this algorithm-centric view gives little attention to the needs of the annotators, especially their needs for transparency~\cite{amershi2014power}. For example, "stopping criteria", knowing when to complete the training with confidence remains a challenge in AL, since the annotator is unable to monitor the model's learning progress. Even if performance metrics calculated on test data are available, it is difficult to judge whether the model will generalize in the real-world context or is bias-free.



Meanwhile, the notion of model transparency has moved beyond the scope of descriptive characteristics of the model studied in prior iML work (e.g., output, performance, features used~\cite{kulesza2015principles,rosenthal2010towards, fails2003interactive,fogarty2008cueflik}). Recent work in the field of explainable AI (XAI)~\cite{gunning2017explainable} focuses on making the \textit{reasoning} of model decisions understandable by people of different roles, including those without formal ML training. In particular, \textit{local explanations} (e.g.~\cite{lundberg2017unified,ribeiro2016should}) is a cluster of XAI techniques that explain how the model arrived at a particular decision. Although researchers have only begun to examine how people actually interact with AI explanations, we believe explanations should be a core component of the interfaces to teach learning models. 

Explanations play a critical role in human teaching and learning~\cite{wellman2004theory,meyer1997consensually}. Prompting students to generate explanations for a given answer or phenomenon is a common teaching strategy to deepen students' understanding. The explanations also enable the teacher to gauge the students' grasp of new concepts, reinforce successful learning, correct misunderstanding, repair gaps, as well as adjust the teaching strategies~\cite{lombrozo2012explanation}. Intuitively, the same mechanism could enable machine teachers to assess the model logic, oversee the machine learner's progress, and establish trust and confidence in the final model. Well-designed explanations could also allow people without ML training to access the inner working of the model and identify its shortcomings, thus potentially reducing the barriers to provide knowledge input and enriching teaching strategies, for example by giving direct feedback for the model's explanations.  

Toward this vision of ``machine teaching through model explanations'', we propose a novel paradigm of \textit{explainable active learning} (XAL), by providing local explanations of the model's predictions of selected instances as the interface to query an annotator's knowledge input. We conduct an empirical study to investigate how local explanations impact the annotation quality and annotator experience. It also serves as an elicitation study to explore how people naturally want to teach a learning model with its explanations. The contributions of this work are threefold:
\begin{itemize}
    \item We provide insights into the opportunities for explainable AI (XAI) techniques as an interface for machine teaching, specifically feature importance based local explanation. We illustrate both the benefits of XAI for machine teaching, including supporting trust calibration and enabling rich teaching feedback, and challenges that future XAI work should tackle, such as anchoring judgment and cognitive workload. We also identify important individual factors mediating one's reception to model explanations in the machine teaching context, including task knowledge, AI experience and Need for Cognition.
    
    \item We conduct an in-depth empirical study of interaction with an active learning algorithm. Our results highlight several problems faced by annotators in an AL setting, such as increasing challenge to provide correct labels as the model matures and selects more uncertain instances, difficulty to know when to stop with confidence, and desire to provide knowledge input beyond labels. We claim that some of these problems can be mitigated by explanations.
    
    \item We propose a new paradigm to teach ML models, \textit{explainable active learning (XAL)}, that has the model selectively query the machine teacher, and meanwhile allows the teacher to understand the model's reasoning and adjust their input. The user study provides a systematic understanding on the feasibility of this new model training paradigm. Based on our findings, we discuss future directions of technical advancement and design opportunities for XAL.
\end{itemize}{}

In the following, we first review related literature, then introduce the proposal for XAL, research questions and hypotheses for the experimental study. Then we discuss the XAL setup, methodology and results. Finally, we reflect on the results and discuss possible future directions.

\section{Related work}
\label{literature}
Our work is motivated by prior work on AL, interactive machine learning and explainable AI. 

\subsection{Active learning}
 The core idea of AL is that if a learning algorithm intelligently selects instances to be labeled, it could perform well with much less training data~\cite{settles2009active}. This idea resonates with the critical challenge in modern ML, that labeled data are time-consuming and expensive to obtain~\cite{zhu2005semi}.  AL can be used in different scenarios like stream based~\cite{cohn1994improving} (from a stream of incoming data),  pool based~\cite{lewis1994sequential} (from a large set of unlabeled instances), etc.~\cite{settles2009active}. To select the next instance for labeling, multiple query sampling strategies have been proposed in the literature \cite{qbc, qbc2, unc, dasgupta2008hierarchical, quire, entropy, confidence}. Most commonly used is \textit{Uncertainty sampling} \cite{unc, entropy, confidence, margin}, which selects instances the model is most uncertain about.  Different AL algorithms exploit different notions of uncertainty, e.g. entropy \cite{entropy}, confidence \cite{confidence}, margin \cite{margin}, etc. 
 

While the original definition of AL is concerned with instance labels, it has been broadened to query other types of knowledge input. Several works explored querying feedback for features, such as asking whether the presence of a feature is an indicator for the target concept~\cite{raghavan2006active,druck2009active,settles2011closing}. For example, DUALIST~\cite{settles2011closing} is an active learning tool that queries annotators for labels of both instances (e.g., whether a text document is about ``baseball'' or ``hockey'') and features (which keywords, if appeared in a document, are likely indicators that the document is about ``baseball''). Other AL paradigms include \textit{active class selection}~\cite{lomasky2007active} and \textit{active feature acquisition}~\cite{zheng2002active}, which query the annotator for additional training examples and missing features, respectively.  

Although AL by definition is an interactive annotation paradigm, the technical ML community tends to simply assume the human annotators to be mechanically queried oracles. The above-mentioned AL algorithms were mostly experimented with simulated human input providing error-free labels. But labeling errors are inevitable, even for simple perceptual judgment tasks~\cite{cheng2015measuring}. Moreover, in reality, the targeted use cases for AL are often ones where high-quality labels are costly to obtain either because of knowledge barriers or effort to label. For example, AL can be used to solicit users' labels for their own records to train an email spam classifier or context-aware sensors ~\cite{kapoor2010interactive,rosenthal2010towards}, but a regular user may lack the knowledge or contextual information to make all judgments correctly. Many have criticized the unrealistic assumptions that AL algorithms make. For example, by solving a multi-instance, multi-oracle optimization problem, \textit{proactive learning}~\cite{donmez2008proactive} relaxes the assumptions that the annotator is infallible, indefatigable (always answers with the same level of quality), individual (only one oracle), and insensitive to costs.

Despite the criticism, we have a very limited understanding on how people actually interact with AL algorithms, hindering our ability to develop AL systems that perform in practice and provide a good annotator experience. Little attention has been given to the annotation interfaces, which in current AL works are undesirably minimal and opaque. To our knowledge, there has been few HCI work on this topic. One exception is in the field of human-robot interaction (HRI), where AL algorithms were used to develop robots that continuously learn by asking humans questions~\cite{cakmak2010designing,cakmak2012designing,chao2010transparent,gonzalez2014asking,saponaro2011generation}. In this context, the robot and its natural-language queries \textit{is} the interface for AL. For example,  Cakmak et al. explored robots that ask three types of AL queries~\cite{cakmak2010designing,cakmak2012designing}: instance queries, feature queries and demonstration queries. The studies found that people were more receptive of feature queries and perceived robots asking about features to be more intelligent. The study also pointed out that a constant stream of queries led to a decline in annotators' situational awareness~\cite{cakmak2010designing}. This kind of empirical results challenged the assumptions made by AL algorithms, and inspired follow-up work proposing mixed-initiative AL: the robot only queries when certain conditions were met, e.g., following an uninformative label. Another relevant study by Rosenthal and Dey ~\cite{rosenthal2010towards} looked at information design for an intelligent agent that queries labels to improve its classification. They found that contextual information, such as keywords in a text document or key features in sensor input, and providing system's prediction (so people only need to confirm or reject labels) improved labeling accuracy. Although this work cited the motivation for AL, the study was conducted with an offline questionnaire without interacting with an actual AL algorithm.

We argue that it is necessary to study annotation interactions with a real-time AL algorithm
because temporal changes are key characteristics of AL settings. With an interactive learning algorithm, every annotation impacts the subsequent model behaviors, and the model should become better aligned with the annotator's knowledge over time. Moreover, systematic changes could happen in the process in both the type of queried instances, depending on the sampling strategy, and the annotator behaviors, for example fatigue~\cite{settles2011closing}. These complex patterns could only be understood by holistically studying the annotation and and the evolving model in real time. 

Lastly, it is a nontrivial issue to understand how annotator characteristics impact their reception to AL system features. For example, it would be instrumental to understand what system features could narrow the performance gaps of people with different levels of domain expertise or AI experience, thus reducing the knowledge barriers to teach ML models.

\subsection{Interactive machine learning}

Active learning is sometimes considered a technique for iML. iML work is primarily motivated by enabling non-ML-experts to train a ML model
through ``rapid, focused, and incremental model updates''~\cite{amershi2014power}. However, conventional AL systems, with a minimum interface asking for labels, lack the fundamental element in iML--a tight interaction loop that transparently presents how every human input impacts the model, so that the non-ML-experts could adapt their input to drive the model into desired directions~\cite{amershi2014power,fails2003interactive}.  Our work aims to move AL in that direction.

Broadly, iML encompasses all kinds of ML tasks including supervised ML, unsupervised ML (e.g., clustering ~\cite{choo2013utopian,smith2018closing}) and reinforcement learning~\cite{cakmak2010designing}. To enable interactivity, iML work has to consider two coupled aspects: \textit{what information} the model presents to people, and \textit{what input} people give to the model. Most iML systems present users with \textit{performance} information as impacted by their input, either performance metrics~\cite{kapoor2010interactive,amershi2015modeltracker}, or model output, for example by visualizing the output for a batch of instances~\cite{fogarty2008cueflik} or allowing users to select instances to inspect. An important lesson from the bulk of iML work is that users value \textit{transparency} beyond performance~\cite{rosenthal2010towards,kulesza2013too}, such as descriptive information about how the algorithm works or what features are used~\cite{kulesza2015principles,rosenthal2010towards}. Transparency is found to not only help improve users' mental model of the learning model and hence provide more effective input, but also satisfaction in their interaction outcomes~\cite{kulesza2013too}. 

iML research has studied a variety of user input into the model such as providing labels, training examples~\cite{fails2003interactive}, as well as specifying model and algorithm choice~\cite{talbot2009ensemblematrix}, parameters, error preferences~\cite{kapoor2010interactive}, etc. A promising direction for iML to out-perform traditional approaches to training ML models is to enable feature-level human input. Intuitively,  direct manipulation of model features represents a much more efficient way to inject domain knowledge into a model~\cite{simard2017machine} than providing labeled instances. For example, FeatureInisght~\cite{brooks2015featureinsight} supports ``feature ideation'' for users to create dictionary features (semantically related groups of words) for text classification. EluciDebug~\cite{kulesza2015principles} allows users to add, remove and adjust the learned weights of keywords for text classifiers. Several interactive topic modeling systems allow users to select keywords or adjust keyword weights for a topic~\cite{choo2013utopian,smith2018closing}. Although the empirical results on whether feature-level input from end users improves performance per se have been mixed~\cite{kulesza2015principles,ahn2007open,wu2019local,stumpf2009interacting}, the consensus is that it is more efficient (i.e., fewer user actions) to achieve comparable results to instance labeling, and that it could produce models better aligned with an individual's needs or knowledge about a domain.

It is worth pointing out that all of the above-mentioned iML and AL systems supporting feature-level input are for text-based models~\cite{settles2011closing,raghavan2006active,stumpf2007toward,smithno,kulesza2015principles}. We suspect that, besides algorithmic interest, the reason is that it is much easier for lay people to consider keywords as top features for text classifiers compared to other types of data. For example, one may come up with keywords that are likely indicators for the topic of ``baseball'', but it is challenging to rank the importance of attributes in a tabular database of job candidates.  One possible solution is to allow people to access the model's own reasoning with features and then make incremental adjustments. This idea underlies recent research into visual analytical tools that support debugging or feature engineering work~\cite{krause2016interacting,hohman2019gamut,wexler2019if}. However, their targeted users are data scientists who would then go back to the model development mode. For non-ML-experts, they would need more accessible information to understand the inner working of the model and provide direct input that does not require heavy work of programming or modeling. Therefore, we propose to leverage recent development in the field of explainable AI as interfaces for non-ML experts to understand and teach learning models.

\subsection{Explainable AI}\label{literature}

The field of explainable AI (XAI))~\cite{gunning2017explainable,guidotti2019survey}, often referred interchangeably as interpretable Machine Learning~\cite{carvalho2019machine,doshi2017towards},  started as a sub-field of AI that aims to produce methods and techniques that make AI's decisions understandable by people. The field has surged in recent years as complex and opaque AI technologies such as deep neural networks are now widely used. Explanations of AI are sought for various reasons, such as by regulators to assess model compliance, or by end users to support their decision-making~\cite{zhang2020effect,liao2020questioning,tomsett2018interpretable}. Most relevant to our work, explanations allow model developers to detect a model's faulty behaviors and evaluate its capability, fairness, and safety~\cite{doshi2017towards,dodge2019explaining}. Explanations are therefore increasingly incorporated in ML development tools supporting debugging tasks such as performance analysis~\cite{ren2016squares}, interactive debugging~\cite{kulesza2015principles}, feature engineering~\cite{krause2014infuse}, instance inspection and model comparison~\cite{hohman2019gamut,zhang2018manifold}.

There have been many recent efforts to categorize the ever-growing collection of explanation techniques~\cite{guidotti2019survey,mohseni2018multidisciplinary,anisi03,lim2019these,wang2019designing,lipton2018mythos,arya2019one}. We focus on those explaining ML classifiers (as opposed to other types of AI system such as planning~\cite{chakraborti2020emerging} or multi-agent systems~\cite{rosenfeld2019explainability}). Guidotti et al. summarized the many forms of explanations as solving three categories of problems: \textit{model explanation} (on the whole logic of the classifier), \textit{outcome explanation} (on the reasons of a decision on a given instance) and \textit{model inspection} (on how the model behaves if changing the input). The first two categories, model and outcome explanations, are also referred as \textit{global} and \textit{local}  explanations~\cite{lipton2018mythos,mohseni2018multidisciplinary,arya2019one}. The HCI community have defined explanation taxonomies based on different types of user needs, often referred as intelligibility types~\cite{lim2009and,lim2019these,liao2020questioning} . Based on Lim and Dey's foundational work~\cite{lim2009and,lim2010toolkit}, intelligibility types can be represented by prototypical user questions to understand the AI, including inputs, outputs, certainty, why, why not, how to, what if and when. A recent work by Liao et al.~\cite{liao2020questioning} attempted to bridge the two streams of work by mapping the user-centered intelligibility types to existing XAI techniques. For example, global explanations answer the question ``\textit{how} does the system make predictions'', local explanations respond to ``\textit{why} is this instance given this prediction'', and model inspection techniques typically addresses \textit{why not}, \textit{what if} and \textit{how to}.

Our work leverages local explanations to accompany AL algorithms' instance queries. Compared to other approaches including example based and rule based explanations~\cite{guidotti2019survey}, \textit{Feature importance}~\cite{ribeiro2016should,guidotti2019survey} is the most popular form of local explanations. It justifies the model's decision for an instance by the instance's important features indicative of the decision (e.g., ``because the patient shows symptoms of sneezing, the model diagnosed him having a cold''). Local feature importance can be generated by different XAI algorithms depending on the underlying model and data. Some algorithms are model-agnostic~\cite{ribeiro2016should,lundberg2017unified}, making them highly desirable and popular techniques. Local importance can be presented to users in different formats~\cite{lipton2018mythos}, such as described in texts~\cite{dodge2019explaining}, or by visualizing the importance values~\cite{poursabzi2018manipulating,cheng2019explaining}. 

While recent studies of XAI often found explanations to improve users' understanding of AI systems~\cite{cheng2019explaining,kocielnik2019will,buccinca2020proxy}, empirical results regarding its impact on users' subjective experience such as trust~\cite{cheng2019explaining,poursabzi2018manipulating,zhang2020effect} and acceptance~\cite{kocielnik2019will} have been mixed. One issue, as some argued~\cite{zhang2020effect}, is that explanation is not meant to enhance trust or satisfaction, but rather to appropriately \textit{calibrate} users' perceptions to the model quality. If the model is under-performing, explanations should work towards exposing the algorithmic limitations; if a model is on par with the expected capability, explanation should help foster confidence and trust. Calibrating trust is especially important for AL settings: if explanations could help the annotator appropriately increase their trust and confidence as the model learns, it could help improve their satisfaction with the teaching outcome and confidently apply stopping criteria (knowing when to stop). Meanwhile, how people react to flawed explanations generated by early-stage, naive models, and changing explanations as the model learns, remain open questions~\cite{smithno}. We will empirically answer these questions by comparing annotation experiences in two snapshots of an AL process: an \textit{early stage} annotation task with the initial model, and a \textit{late stage} when the model is close to the stopping criteria.

On the flip side, explanations present additional information and the risk of overloading users~\cite{narayanan2018humans}, although some showed that their benefit justifies the additional effort~\cite{kulesza2015principles}. Explanations were also found to incur over-reliance~\cite{stumpf2016explanations,poursabzi2018manipulating} which makes people less inclined or able to scrutinize AI system's errors. It is possible that explanations could bias, or \textit{anchor} annotators' judgment to the model's. While anchoring judgment is not necessarily counter-productive if the model predictions are competent,  we recognize that the most popular sampling strategy of AL--uncertainty sampling--focuses on instances the model is most uncertain of.  To test this, it is necessary to decouple the potential anchoring effect of the model's predictions~\cite{rosenthal2010towards}, and the model's explanations, as an XAL setting entails both. Therefore, we compare the model training results with XAL to two baseline conditions: traditional AL and \textit{coactive learning} (CL)~\cite{shivaswamy2015coactive}. CL is a sub-paradigm of AL, in which the model presents its predictions and the annotator is only required to make corrections if necessary. CL is favored for reducing annotator workload, especially when their availability is limited. 

Last but not least, recent XAI work emphasizes that there is no ``one-fits-all'' solution and different user groups may react to AI explanations differently~\cite{arya2019one,liao2020questioning,dodge2019explaining}. Identifying individual factors that mediate the effect of AI explanation could help develop more robust insights to guide the design of explanations.
Our study provides an opportunity to identify key individual factors that mediate the preferences for model explanations in the machine teaching context. Specifically, we study the effect of \textit{Task (domain) Knowledge} and \textit{AI Experience} to test the possibilities of XAL for reducing knowledge barriers to train ML models. We also explore the effect of \textit{Need for cognition}~\cite{cacioppo1982need}, defined as an individual's tendency to engage in thinking or complex cognitive activities. Need for cognition has been extensively researched in social and cognitive psychology as a mediating factor for how one responds to cognitively demanding tasks (e.g.~\cite{cacioppo1983effects,haugtvedt1992personality}). Given that explanations present additional information, we hypothesize that individuals with different levels of Need for Cognition could have different responses.

\section{Explainable Active Learning and Research Questions}

We propose \textit{explainable active learning (XAL)} by combining active learning and \textit{local explanations}, which fits naturally with the AL workflow without requiring additional user input:  instead of opaquely requesting instance labels, the model presents its own decision accompanied by its explanation for the decision, answering the question ``\textit{why} am I giving this instance this prediction". It then requests the annotator to confirm or reject. For the user study, we make the design choice of explaining AL with \textit{local feature importance} instead of other forms of local explanations (e.g., example or rule based explanations~\cite{guidotti2019survey}), given the former approach's popularity and intuitiveness--it reflects how the model weighs different features and gives people direct access to the inner working of the model. We also make the design choice of presenting local feature importance with a visualization (Figure~\ref{fig:interface_2}) instead of in texts, in the hope of reading efficiency.

Our idea differentiates from prior work on feature-querying AL and iML in two aspects. First, we present the model's own reasoning for a particular instance to query user feedback instead of requesting global feature weights from people~\cite{settles2011closing,raghavan2006active,kulesza2015principles,brooks2015featureinsight}. Recent work demonstrated that, while ML experts may be able to reason with model features globally, lay people prefer local explanations grounded in specific cases~\cite{arya2019one,kulesza2013too,hohman2019gamut,kulesza2011oriented}. Second, we look beyond text-based models as in existing work as discussed above, and consider a generalizable form of explanation--visualizing local feature importance. While we study XAL in a setting of tabular data, this explanation format can be applied to any type of data with model-agnostic explanation techniques (e.g.~\cite{ribeiro2016should}).

At a high level, we posit that this paradigm of presenting explanations and requesting feedback better mimics how humans teach and learn, allowing transparency for the annotation experience. Explanations can also potentially improve the teaching quality in two ways. First, it is possible that explanations make it easier for one to reject a faulty model decision and thus provide better labels, especially for challenging situations where the annotator lacks contextual information or complete domain knowledge~\cite{rosenthal2010towards}. Second, explanations could enable new forms of teaching feedback based on the explanation. These benefits were discussed in a very recent paper by Teso and Kersting~\cite{teso2018should}, which explored soliciting corrections for the model's explanation, specifically feedback that a mentioned feature should be considered irrelevant instead. This correction feedback is then used to generate counter examples as additional training data, which are identical to the instance except for the mentioned feature. While this work is closest to our idea, empirical studies were absent to understand how adding explanations impacts AL interactions. 

We believe a user study is necessary for two reasons. First, accumulating evidence, as reviewed in the previous section, suggests that explanations have both benefits and drawbacks relevant to an AL setting. They merit a user study to test its feasibility. Second, a design principle of iML recommends that algorithmic advancement should be driven by people's natural tendency to interact with models~\cite{amershi2014power,cakmak2012designing,stumpf2009interacting}. Instead of fixing on a type of input as in Teso and Kersting~\cite{teso2018should}, an \textit{interaction elicitation study} could map out desired interactions for people to teach models based on its explanations and then inform algorithms that are able to take advantage of these interactions. A notable work by Stumpf et al.~\cite{stumpf2009interacting} conducted an elicitation study for interactively improving text-based models, and developed new training algorithms for Naïve Bayes models. Our study explores how people naturally want to teach a model with a local-feature-importance visualization, a popular and generalizable form of explanation. Based on the above discussions, this paper sets out to answer the following research questions and test the following hypotheses:

 

\begin{itemize}
    \item \textbf{RQ1}: How do local explanations impact the annotation and training outcomes of AL?
    \item \textbf{RQ2}: How do local explanations impact annotator experiences?
    
    \begin{itemize}
        \item \textbf{H1}: Explanations support \textit{trust calibration}, i.e. there is an interactive effect between the presence of explanations and the model learning stage (early v.s. late stage model) on annotator's trust in deploying the model. 
        \item \textbf{H2}: Explanations improve \textit{annotator satisfaction}.
        \item \textbf{H3}: Explanations increase perceived \textit{cognitive workload}. 
        
    \end{itemize}{}
 \item \textbf{RQ3}: How do individual factors, specifically \textit{task knowledge}, \textit{AI experience}, and \textit{Need for Cognition}, impact annotation and annotator experiences with XAL?
 
 \begin{itemize}
        \item \textbf{H4}: Annotators with lower task knowledge benefit more from XAL, i.e., there is an interactive effect between the presence of explanations and annotators' task knowledge on some of the annotation outcome and experience measures  (trust, satisfaction or cognitive workload). 
        
        \item \textbf{H5}:  Annotators inexperienced with AI benefit more from XAL, i.e., there is an interactive effect between the presence of explanations and annotators' experience with AI on some of the annotation outcome and experience measures (trust, satisfaction or cognitive workload).
        
        \item \textbf{H6}: Annotators with lower Need for Cognition have a less positive experience with XAL, i.e., there is an interactive effect between the presence of explanations and annotators' Need for Cognition on some of the annotation outcome and experience measures (trust, satisfaction or cognitive workload),

    \end{itemize}{}
 \item \textbf{RQ4}: What kind of feedback do annotators naturally want to provide upon seeing local explanations?
\end{itemize}{}

\section{XAL Setup}

\subsection{Prediction task}
We aimed to design a prediction task that would not require deep domain expertise, where common-sense knowledge could be effective for teaching the model. The task should also involve decisions by weighing different features so explanations could potentially make a difference (i.e., not simple perception based judgment). Lastly, the instances should be easy to comprehend with a reasonable number of features. With these criteria, we chose the Adult Income dataset \cite{adultIncome} for a task of predicting whether the annual income of an individual is more or less than \$80,000~\footnote{After adjusting for inflation (1994-2019)~\cite{inflation}, while the original dataset reported on the income level of \$50,000}. The dataset is based on a Census survey database. Each row in the dataset characterizes a person with a mix of numerical and categorical variables like age, gender, education, occupation, etc., and a binary annual income variable, which was used as our ground truth.

\begin{figure*}[ht]
    \centering
\begin{subfigure}{.47\textwidth}
  \centering
  \includegraphics[width=1.1\linewidth]{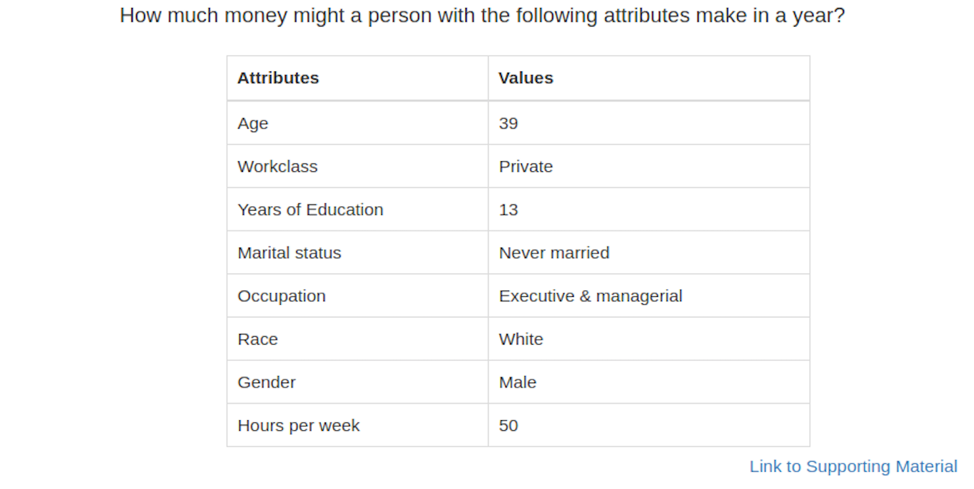}
  \caption{Customer profile presented in all conditions for annotation}
  \label{fig:interface_1}
  \label{fig:sub1}
\end{subfigure} \hspace{5mm}
\begin{subfigure}{.47\textwidth}
  \centering
  \includegraphics[width=.8\linewidth]{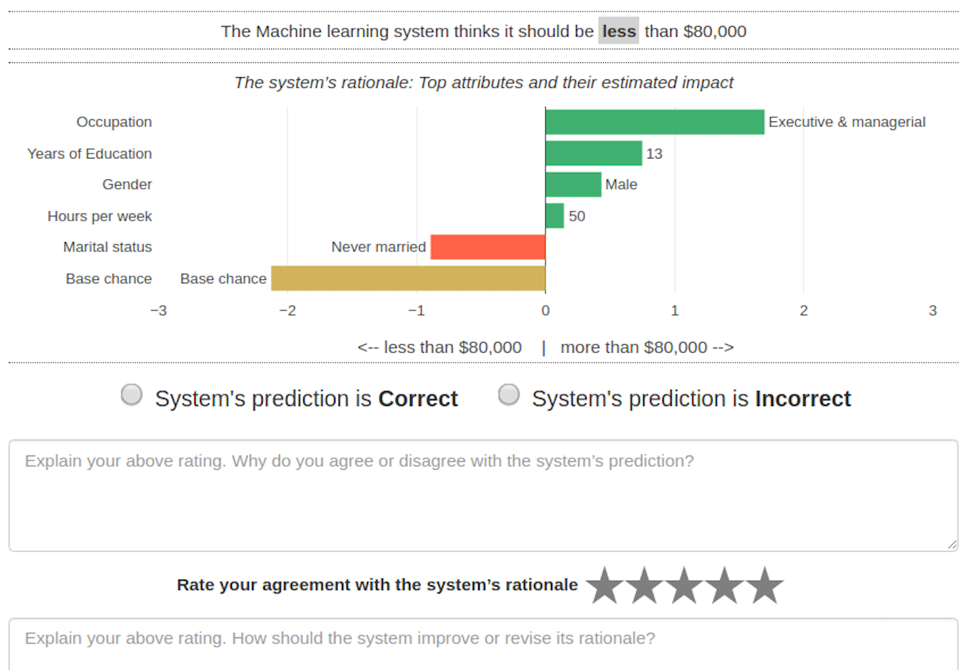}
  \caption{Explanation and questions presented in the XAL condition}
  \label{fig:interface_2}
  \label{fig:sub2}
\end{subfigure}
    \caption{Experiment interface }
    \label{fig:interface}
\end{figure*}

In the experiment, we presented participants with a scenario of building an ML classification system for a customer database. Based on a customer's background information, the system predicts the customer's income level for a targeted service. The task for the participants was to judge the income level of instances that the system selected to learn from, as presented in Figure~\ref{fig:interface_1}. This is a realistic AL task where annotators might not provide error-free labels, and explanations could potentially help reveal faulty model beliefs. To improve participants' knowledge about the domain, we provided a practice task before the trials, which will be discussed in  Section~\ref{domain}.

\subsection{Active learning setup}

AL requires the model to be retrained after new labels are fetched, so the model and explanations used for the experiment should be computationally inexpensive to avoid latency. Therefore we chose logistic regression (with L2 regularization), which was used extensively in the AL literature \cite{settles2009active, yang2018benchmark}. Logistic regression is considered directly interpretable, i.e., its local feature importance could be directly generated, as to be described in Section~\ref{explanation}. We note that this form of explanation could be generated by post-hoc techniques for any kind of ML model~\cite{ribeiro2016should}.

\begin{figure}
    \centering
    \includegraphics[scale=0.5]{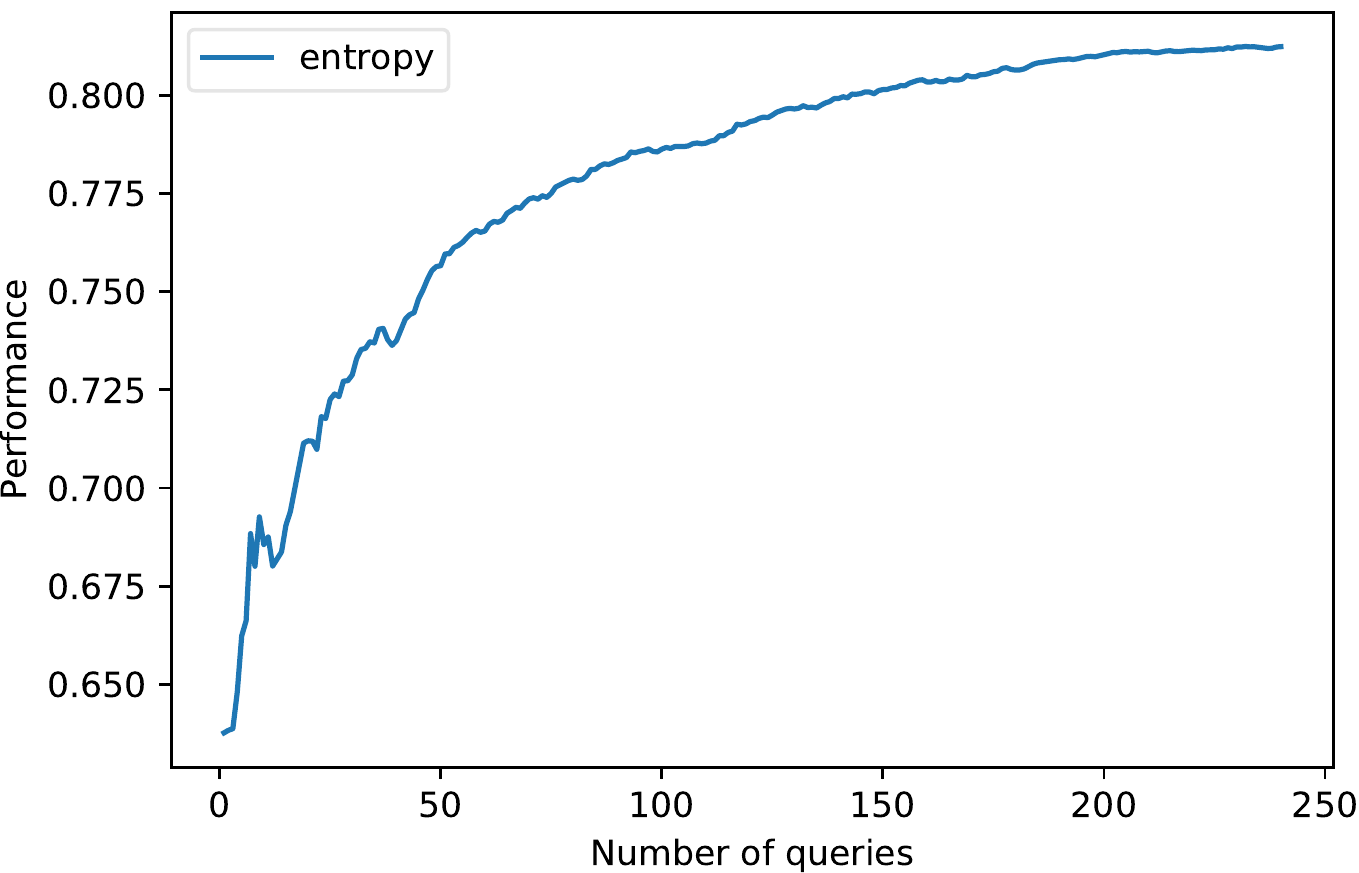}
    \caption{Accuracy as a function of number of queries in the simulation experiment}
    \label{fig:accuracy}
\end{figure}

Building an AL pipeline involves the design choices of sampling strategy, batch size, the number of initial labeled instances and test data. For this study, we used entropy-based uncertainty sampling to select the next instance to query, as it is the most commonly used sampling strategy \cite{yang2018benchmark} and also computationally inexpensive. We used a batch size of 1~\cite{batchSize}, meaning the model was retrained after each new queried label. We initialized the AL pipeline with two labeled instances. To avoid tying the experiment results to a particular sequence of data, we allocated different sets of initial instances to different participants, by randomly drawing from a pool of more than 100 pairs of labeled instances. The pool was created by randomly picking two instances with ground-truth labels, and being kept in the pool only if they produced a model with initial accuracy between 50\%-55\%. This was to ensure that the initial model would perform worse than humans and did not vary significantly across participants. 25\% of all data were reserved as test data for evaluating the model learning outcomes.

As discussed, we are interested in the effect of explanations at different stages of AL. We took two snapshots of an AL process--an early-stage model just started with the initial labeled instances, and a late-stage model that is close to the stopping criteria. We define the stopping criteria as plateau of accuracy improvement on the test data with more labeled data. To determine where to take the late-stage snapshot, we ran a simulation where AL queried instances were given the labels in the ground truth. The simulation was run with 10 sets of initial labels and the mean accuracy is shown in Figure \ref{fig:accuracy}. Based on the pattern, we chose the late stage model to be where 200 queries were executed. To create the late-stage experience without having participants answer 200 queries, we took a participant's allocated initial labeled instances and simulated an AL process with 200 queries answered by the ground-truth labels. The model was then used in the late-stage task for the same participant. This also ensured that the two tasks a participant experienced were independent of each other i.e. a participant's performance in the early-stage task did not influence the late-stage task. In each task, participants were queried for 20 instances. Based on the simulation result in Figure~\ref{fig:accuracy}, we expected an improvement of 10\%-20\% accuracy with 20 queries in the early stage, and a much smaller increase in the late stage.

\subsubsection{Explanation method}\label{explanation}
Figure~\ref{fig:interface_2} shows a screenshot of the local explanation presented in the XAL condition, for the instance shown in Figure~\ref{fig:sub1}. The explanation was generated based on the coefficients of the logistic regression, which determine the impact of each feature on the model's prediction. To obtain the \textit{feature importance} for a given instance, we computed the product of each of the instance's feature values with the corresponding coefficients in the model. The higher the magnitude of a feature's importance, the more impact it had on the model's prediction for this instance. A negative value implied that the feature value was tilting the model's prediction towards less than \$80,000 and vice versa. We sorted all features by their absolute importance and picked the top 5 features responsible for the model's prediction. 

The selected features were shown to the participants in the form of a horizontal bar chart as in Figure~\ref{fig:interface_2}. The importance of a feature was encoded by the length of the bar where a longer bar meant greater impact and vice versa. The sign of the feature importance was encoded with color (green-positive, red-negative), and sorted to have the positive features at the top of the chart. Apart from the top contributing features, we also displayed the intercept of the logistic regression model as an orange bar at the bottom. Because it was a relatively skewed classification task (the majority of the population has an annual income of less than \$80,000), the negative base chance (intercept) needed to be understood for the model's decision logic. For example, in Figure \ref{fig:interface}, Occupation is the most important feature. Martial status and base chance are pointing towards less than \$80,000. While most features are tilting positively, the model prediction for this instance is still less than \$80,000 because of the large negative value of base chance.   



\section{Experimental design}

We adopted a 3 $\times$ 2 experimental design, with the learning condition (AL, CL, XAL) as a between-subject treatment, and the learning stage (early v.s. late) as a within-subject treatment. That is, participants were randomly assigned to one of the conditions to complete two tasks, with queries from an early and a late stage AL model, respectively. The order of the early and late stage tasks was randomized and balanced for each participant to avoid order effect and biases from knowing which was the "improved" model.

We posted the experiment as a human intelligence task (HIT) on Amazon Mechanical Turk. We set the requirement to have at least 98\% prior approval rate and each worker could participate
only once. Upon accepting the HIT, a participant was assigned to one of the three conditions. The annotation task was given with a scenario of building a classification system for a customer database to provide targeted service for high- versus low-income customers, with a ML model that queries and learns in real time. Given that the order of the learning stage was randomized, we instructed the participants that they would be teaching two configurations of the system with different initial performance and learning capabilities. 

With each configuration, a participant was queried for 20 instances, in the format shown in Figure~\ref{fig:interface_1}. A minimum of 10 seconds was enforced before they could proceed to the next query. In the AL condition, participants were presented with a customer's profile and asked to judge whether his or her annual income was above 80K. In the CL condition, participants were presented with the profile and the model's prediction. In the XAL condition, the model's prediction was accompanied by an explanation revealing the model's "rationale for making the prediction" (the top part of Figure~\ref{fig:interface_2}). In both the CL and XAL conditions, participants were asked to judge whether the model prediction was correct and optionally answer an open-form question to explain that judgement (the middle part of Figure~\ref{fig:interface_2}). In the XAL condition, participants were further asked to also give a rating to the model explanation and optionally explain their ratings with an open-form question (the bottom part of Figure~\ref{fig:interface_2}).  After a participant submitted a query, the model was retrained, and performance metrics of accuracy and F1 score (on the 25\% reserved test data) were calculated and recorded, together with the participant's input and the time stamp.
 
After every 10 trials, the participants were told the percentage of their answers matching similar cases in the Census survey data, as a measure to help engaging the participants. An attention-check question was prompted in each learning stage task, showing the customer's profile in the prior query with two other randomly selected profiles as distractors. The participants were asked to select the one they just saw. Only one participant failed both attention-check questions, and was excluded from the analysis. 

After completing 20 queries for each learning stage task, the participants were asked to fill out a survey regarding their subjective perception of the ML model they just finished teaching and the annotation task. The details of the survey will be discussed in Section ~\ref{survey}. At the end of the HIT we also collected participants' demographic information and factors of individual differences, to be discussed in Section~\ref{individual}.

\subsubsection{Domain knowledge training} \label{domain}
We acknowledge that MTurk workers may not be experts of an income prediction task, even though it is a common topic. Our study is close to \textit{human-grounded evaluation} proposed in ~\cite{doshi2017towards} as an evaluation approach for explainability, in which lay people are used as proxy to test general notions or patterns of the target application (i.e., by comparing outcomes between the baseline and the target treatment). 

To improve the external validity, we took two measures to help participants gain domain knowledge. First, throughout the study, we provided a link to a supporting document with statistics of personal income based on the Census survey. Specifically, chance numbers--the chance of people with a feature-value to have income above 80K--were given for all feature-values the model used (by quantile if numerical features). Second, participants were given 20 practice trials of income prediction tasks and encouraged to utilize the supporting material. The ground truth--income level reported in the Census survey--was revealed after they completed each practice trial. Participants were told that the model would be evaluated based on data in the Census survey, so they should strive to bring the knowledge from the supporting material and the practice trials into the annotation task. They were also incentivized with a \$2 bonus if the consistency between their predictions and similar cases reported in the Census survey were among the top 10\% of all participants.

After the practice trials, the agreement of the  participants' predictions with the ground-truth in the Census survey for the early-stage trials reached a mean of 0.65 (SE=0.08). We note the queried instances in AL using uncertainty-based sampling are challenging by nature. The agreement with ground truth by one of the authors, who is highly familiar with the data and the task, was 0.75.

\subsubsection{Survey measuring subjective experience}\label{survey}
To understand how explanation impacts annotators' subjective experiences (\textbf{RQ2}), we designed a survey for the participants to fill after completing each learning stage task. We asked the participants to self report the following (all based on a 5-point Likert Scale):

\textit{Trust} in deploying the model: We asked participants to assess how much they could trust the model they just finished teaching to be deployed for the target task (customer classification). Trust in technologies is frequently measured based on McKnight’s framework on Trust~\cite{mcknight1998initial,mcknight2002developing}, which considers the dimensions of \textit{capability}, \textit{benevolence}, \textit{integrity} for trust belief, and multiple action-based items (e.g., "I will be able to rely on the system for the target task") for trust intention. We also consulted a recent paper on trust scale for automation~\cite{korber2018theoretical} and added the dimension of \textit{predictability} for trust belief. We picked and adapted one item in each of the four trust belief dimensions (e.g., for benevolence, "Using predictions made by the system will harm customers’ interest") , and four items for trust intention, and arrived at an 8-item scale to measure trust (3 were reversed scale).  The Cronbach's alpha is 0.89.

\textit{Satisfaction} of the annotation experience, by five items adapted from After-Scenario Questionnaire~\cite{lewis1995computer} and User Engagement Scale~\cite{o2018practical} (e.g. "I am satisfied with the ease of completing the task", "It was an engaging experience working on the task"). The Cronbach's alpha is 0.91

\textit{Cognitive workload} of the annotation experience, by selecting two applicable items from the NASA-TLX task load index (e.g., "How mentally demanding was the task: 1=very low; 5=very high"). The Cronbach's alpha is 0.86.


\subsubsection{Individual differences}\label{individual}
\textbf{RQ3} asks about the mediating effect of individual differences, specifically the following:

\textit{Task knowledge} to perform the income prediction judgement correctly. We used one's performance in the practice trails as a proxy, calculated by the percentage of trials judged correctly based on the ground truth of income level in the Census database.

\textit{AI experience}, for which we asked participants to self-report ``How much do you know about artificial Intelligence or machine learning algorithms.'' The original questions had four levels of experience. With few answered higher level of experience, we decided to combine the answers into a binary variable--without AI experience v.s. with AI experience.

\textit{Need for Cognition} measures individual differences in the tendency to engage in thinking and cognitively complex activities. To keep the survey short, we selected two items from the classic Need for Cognition scale developed by Cacioppo and Petty~\cite{cacioppo1982need}. The Cronbach's alpha is 0.88.

\subsubsection{Participants}
37 participants completed the study. One participant did not pass both attention-check tests and was excluded. The analysis was conducted with 12 participants in each condition. Among them, 27.8\% were female; 19.4\% under the age 30, and 13.9\% above the age 50; 30.6\% reported to have no knowledge of AI, 52.8\% with little knowledge ("know basic concepts in AI"), and the rest to have some knowledge ("know or used AI algorithms").  In total , participants spent about 20-40 min on the study and was compensated for \$4 with a 10\% chance for additional \$2 bonus, as discussed in Section~\ref{domain}

\section{Results}
 For all analyses, we ran mixed-effects regression models to test the hypotheses and answer the research questions, with participants as random effects,
 learning \textit{Stage}, \textit{Condition}, and individual factors (\textit{Task Knowledge}, \textit{AI Experience}, and \textit{Need for Cognition}) as fixed effects. RQ2 and RQ3 are concerned with interactive effects of Stage or Individual factors with learning Conditions. Therefore for every dependant variable we are interested in, we started with including all two-way interactions with Condition in the model, then removed insignificant interactive terms in reducing order. A VIF test was run to confirm there was no multicollinearity issue with any of the variables (all lower than 2). In each sub-section, we report statistics based on the final model and summarize the findings at the end.

\subsection{Annotation and learning outcomes (\textbf{RQ1}, \textbf{RQ3})}

First, we examined the model learning outcomes in different conditions. In Table~\ref{tab:performance} (the third to sixth columns), we report the statistics of performance metrics--\textit{Accuracy} and \textit{F1} scores-- after the 20 queries in each condition and learning stage. We also report the performance improvement, as compared to the initial model performance before the 20 queries.

For each of the performance and improvement metrics, we ran a mixed-effect regression model as described earlier. In all the models, we found only significant main effect of Stage for all performance and improvement metrics ($p<0.001$). The results indicate that participants were able to improve the early-stage model significantly more than the later-stage model, but the improvement did not differ across learning conditions. 


\begin{table}
  \caption{Results of model performance and labels }\label{tab:performance}
   
  \begin{tabular}{p{1cm}p{1.2cm}p{1.4cm}p{1.4cm}p{1.4cm}p{1.4cm}p{1.4cm}p{1.4cm}}
    \toprule
    Stage&Condition&Acc.&Acc. improve&F1&F1 improve&\%Agree&Human Acc.\\
    \midrule
    &AL & 67.0\% & 13.7\% & 0.490 & 0.104 & 55.0\% & 66.7\%\\
    Early &CL & 64.2\% & 11.7\% & 0.484 & 0.105 & 58.3\% & 62.1\%\\
    &XAL & 64.0\% & 11.8\% & 0.475 & 0.093 & 62.9\% & 63.3\%\\
    \midrule
    &AL & 80.4\% & 0.1\% & 0.589 & 0.005 & 47.9\% & 54.2\%\\
    Late &CL & 80.8\% & 0.2\% & 0.587 & 0.007 & 55.8\% & 58.8\%\\
    &XAL & 80.3\% & -0.2\% & 0.585 & -0.001 & 60.0\% & 55.0\%\\
  \bottomrule
\end{tabular}

\end{table}

\begin{figure}
    \centering
    \includegraphics[width=0.7\linewidth]{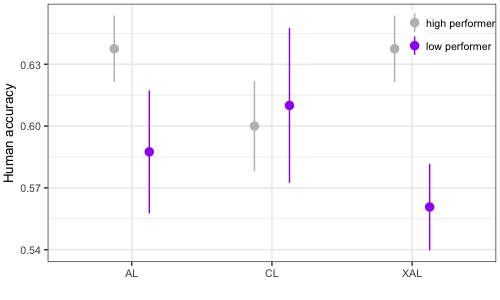}
    \caption{Human accuracy across conditions and task knowledge levels. All error bars represent +/- one standard error.}
    \label{fig:human_acc}
\end{figure}

In addition to the performance metrics, we looked at the \textit{Human accuracy}, defined as the percentage of labels given by a participant that were consistent with the ground truth. Interestingly, we found a significant interactive effect between Condition and participants' Task Knowledge (calculated as one's accuracy score in the training trials): taking CL condition as a reference level, XAL had a positive interactive effect with Task Knowledge ($\beta=0.67,SE=0.29, p=0.03$). In Figure~\ref{fig:human_acc}, we plot the pattern of the interactive effect by first performing a median split on Task Knowledge scores to categorize participants into \textit{high performers} and \textit{low performers}. The figure shows that, compared to the CL condition, adding explanations had a reverse effect for those with high or low task knowledge. While explanations helped those with high task knowledge to provide better labels, it impaired the judgment of those with low task knowledge. There was also a main negative effect of late Stage ($SE=0.21,t=3.87,p<0.001$), confirming that queried instances in the later stage were more challenging for participants to judge correctly. 


\begin{figure}
    \centering
    \includegraphics[width=0.7\linewidth]{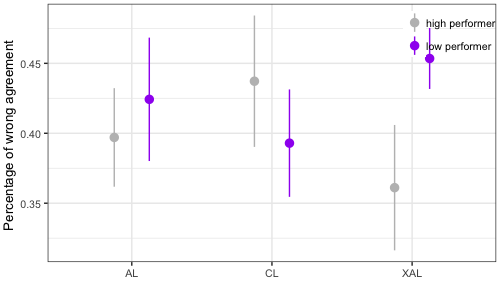}
    \caption{Percentage of wrong agreement among all agreeing judgments with model predictions across Conditions and Task Knowledge levels. All error bars represent +/- one standard error.}
    \label{fig:wrong_agree}
\end{figure}

We conducted the same analysis on the \textit{Agreement} between each participant's labels and the model predictions and found a similar trend: using the CL condition as the reference level, there was a marginally significant interactive effect between XAL and Task Knowledge ($\beta=-0.75, SE=0.45,p=0.10$) \footnote{We consider $p<0.05$ as significant, and $0.05 \leq p<0.10$ as marginally significant, following statistical convention~\cite{cramer2004sage}}. The result suggests that explanations might have an "anchoring effect" on those with low task knowledge, making them more inclined to accept the model's predictions. Indeed, we zoomed in on trials where participants agreed with the model predictions, and looked at the percentage of \textit{wrong agreement} where the judgment was inconsistent with the ground truth. We found a significant interaction between XAL and Task Knowledge, using CL as a reference level ($\beta=-0.89, SE=0.45,p=0.05$). We plot this interactive effect in Figure~\ref{fig:wrong_agree}: adding explanations had a reverse effect for those with high or low task knowledge, making the latter more inclined to mistakenly agree with the model's predictions.  We did not find such an effect for \textit{incorrect disagreement} looking at trials where participants disagreed with the model's predictions.

Taken together, to our surprise, we found the opposite results of \textbf{H4}: local explanations further polarized the annotation outcomes of those with high or low task knowledge, compared to only showing model predictions without explanations. While explanations may help those with high task knowledge to make better judgment, they have a \textbf{negative anchoring effect for those with low task knowledge by making them more inclined to agree with the model even if it is erroneous}. This could be a potential problem for XAL, even though we did not find this anchoring effect to have statistically significant negative impact on the model's learning outcome. We also showed that with uncertainty sampling of AL, \textbf{as the model matured, it became more challenging for annotators to make correct judgment and improve the model performance}.

\subsection{Annotator experience (\textbf{RQ2}, \textbf{RQ3})}
We then investigated how participants' self-reported experience differed across conditions by analyzing the following survey scales (measurements discussed in Section ~\ref{survey}): trust in deploying the model, interaction satisfaction, and perceived cognitive workload. Table~\ref{tab:survey} reports the mean ratings in different conditions and learning stage tasks. For each self-reported scale, we ran a mixed-effects regression model as discussed in the beginning of this section.

\begin{table}
  \caption{Survey results }\label{tab:survey}

  \begin{tabular}{p{0.8cm}p{1.2cm}p{1.5cm}p{1.5cm}p{1.5cm}}
    \toprule
        Stage&Condition&Trust&Satisfaction&Workload\\
    \midrule
    &AL &3.14 &4.23  &2.08\\
    Early&CL &3.83 &3.69 &2.71 \\
    &XAL &2.42 &3.31 &3.00 \\
    \midrule
    &AL &3 & 4.18&2.25\\
    Late&CL &2.71 &3.63 &2.67\\
    &XAL &2.99 &3.35&3.14  \\

  \bottomrule
\end{tabular}

\end{table}

\begin{figure}
    \centering
    \includegraphics[width=0.7\linewidth]{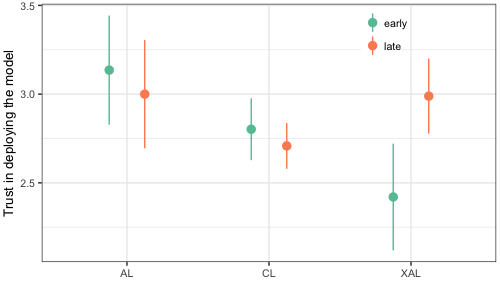}
    \caption{Trust in deploying the model across Conditions and Stages. All error bars represent +/- one standard error.}

    \label{fig:trust_stage}
\end{figure}

First, for trust in deploying the model, using AL as the reference level, we found a significant positive interaction between XAL Condition and Stage ($\beta=0.70, SE=0.31,p=0.03$). As shown in Table~\ref{tab:survey} and Figure~\ref{fig:trust_stage}, compared to the other two conditions, participants in the XAL Condition had significantly lower trust in deploying the early stage model, but enhanced their trust in the later stage model. The results confirmed \textbf{H1 }that \textbf{explanations  help calibrate annotators' trust} in the model at different stages of the training process, while showing model predictions alone (CL) was not able to have that effect.

\begin{figure}
    \centering
    \includegraphics[width=0.7\linewidth]{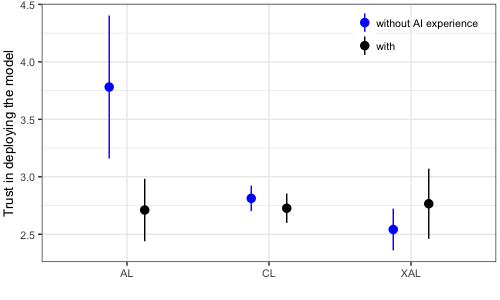}
    \caption{Trust in deploying the model across conditions and experience with AI. All error bars represent +/- one standard error.}
    \label{fig:trust_AI}
\end{figure}

We also found a two-way interaction between XAL Condition and participants' AI Experience (with/without experience) on trust in deploying the model ($\beta=1.43, SE=0.72,p=0.05$) (AL as the reference level). Figure ~\ref{fig:trust_AI} plots the effect: people without AI experience had exceptionally high ``blind'' trust and high variance of the trust (error bar) in deploying the model in the AL condition. With XAL they were able to an appropriate level of trust. The result highlight the \textbf{challenge for annotators to assess the trustworthiness of the model to be deployed, especially for those inexperienced with AI. Providing explanations could effectively appropriate their trust}, supporting \textbf{H5}.

\begin{figure}
    \centering
    \includegraphics[width=0.7\linewidth]{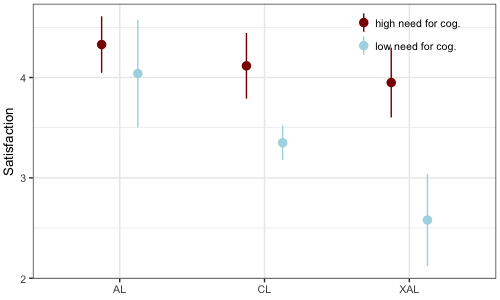}
    \caption{Satisfaction across conditions and experience with AI. All error bars represent +/- one standard error.}
    \label{fig:satisfaction_nc}
\end{figure}

For interaction satisfaction, the descriptive results in Table~\ref{tab:survey} suggests a decreasing trend of satisfaction in XAL condition compared to baseline AL. By running the regression model we found a significant two-way interaction between XAL Condition and Need for Cognition ($\beta=0.54, SE=0.26, p=0.05$) (AL as reference level). Figure ~\ref{fig:satisfaction_nc} plots the interactive effect, with median split on Need for Cognition scores. It demonstrates that \textbf{explanations negatively impacted satisfaction, but only for those with low Need for Cognition}, supporting \textbf{H6} and rejecting \textbf{H2}. We also found a positive main effect of Task Knowledge ($SE=1.31,t=2.76,p=0.01$), indicating that people who were good at the annotation task reported higher satisfaction.


\begin{figure}
    \centering
    \includegraphics[width=0.7\linewidth]{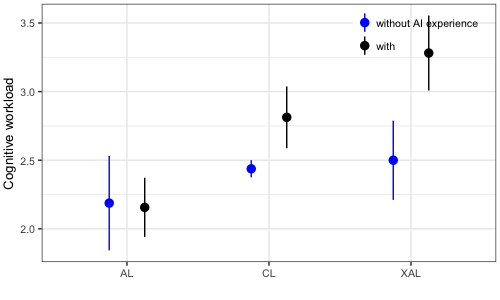}
    \caption{Cognitive workload across conditions and experience with AI. All error bars represent +/- one standard error.}
    \label{fig:workload_AI}
\end{figure}

For self-reported cognitive workload, the descriptive results in Table~\ref{tab:survey} suggests an increasing trend in XAL condition compared to baseline AL. Regression model found an interactive effect between the condition XAL and AI experience ($\beta=1.30, SE=0.59,p=0.04$). As plotted in Figure~\ref{fig:workload_AI}, the \textbf{XAL condition presented higher cognitive workload compared to baseline AL, but only for those with AI experience}. This partially supports \textbf{H3}, and potentially suggests that those with AI experience were able to more carefully examine the explanations. 

We also found an interactive effect between CL condition and Need for Cognition on cognitive workload ($\beta=0.53, SE=0.19,p=0.01$), and the remaining negative main effect of Need for Cognition ($\beta=-0.41, SE=0.14,p=0.01$). Pair-wise comparison suggests that participants with low Need for Cognition reported higher cognitive workload than those with high Need for Cognition, except in the CL condition, where they only had to accept or reject the model's predictions. Together with the results on satisfaction, \textbf{CL may be a preferred choice for those with low Need for Cognition}.


In summary, to answer \textbf{RQ2}, participants' self-reported experience confirmed the benefit of explanations for calibrating trust and judging the maturity of the model. Hence XAL could potentially help annotators form stopping criteria with more confidence. Evidence was found that explanations increased cognitive workload, but only for those experienced with AI. We also identified an unexpected effect of explanations in reducing annotator satisfaction, but only for those self-identified to have low Need for Cognition, suggesting that the additional information and workload of explanation may avert annotators who have little interest or capacity to deliberate on the explanations.

 The quantitative results with regard to \textbf{RQ3} confirmed the mediating effect of individual differences in Task Knowledge, AI Experience and Need for Cognition on one's reception to explanations in an AL setting. Specifically, people with better Task Knowledge and thus more capable of detecting AI's faulty reasoning, people inexperienced with AI who might be otherwise clueless about the model training task, and people with high Need for Cognition, may benefit more from XAL compared to traditional AL.

   



\subsection{Feedback for explanation (\textbf{RQ4})}
In the XAL condition, participants were asked to rate the system's rationale based on the explanations and respond to an optional question to explain their ratings. Analyzing answers to these questions allowed us to understand what kind of feedback participants naturally wanted to give the explanations (\textbf{RQ4}).

\begin{figure}
    \centering
    \includegraphics[scale=0.9]{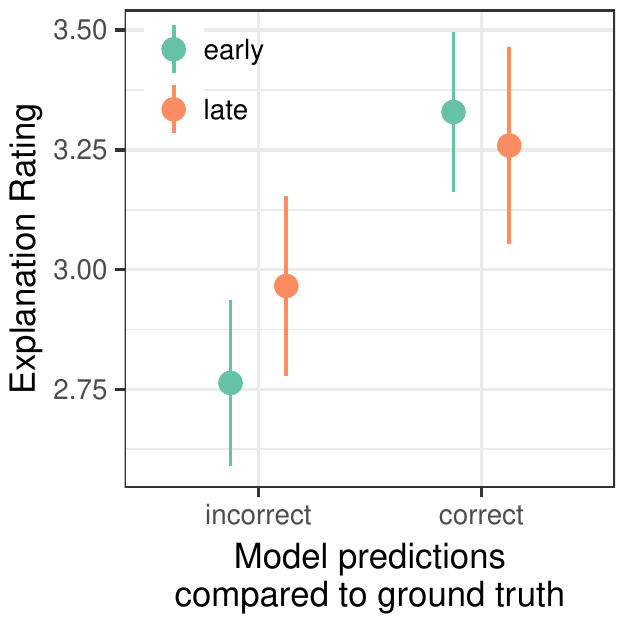}
    \caption{Explanation ratings for correct and incorrect model predictions}
    \label{fig:model_pred}
\end{figure}

First, we inspected whether participants' explanation ratings could provide useful information for the model to learn from. Specifically, if the ratings could distinguish between correct and incorrect model predictions, then they could provide additional signals. Focusing on the XAL condition, we calculated, for each participant, in each learning stage task, the \textit{average explanation ratings} given to instances where the model made correct and incorrect predictions (compared to ground truth). The results are shown in Figure~\ref{fig:model_pred}. By running an ANOVA on the \textit{average explanation ratings}, with \textit{Stage} and \textit{Model Correctness} as within-subject variables, we found the main effect of \textit{Model Correctness} to be significant, $F(1, 11)=14.38$, $p<0.01$. This result indicates that participants were able to distinguish the rationales of correct and incorrect model predictions, in both the early and late stages, confirming the utility of annotators' rating on the explanations.

\begin{figure}
    \centering
    \includegraphics[scale=0.9]{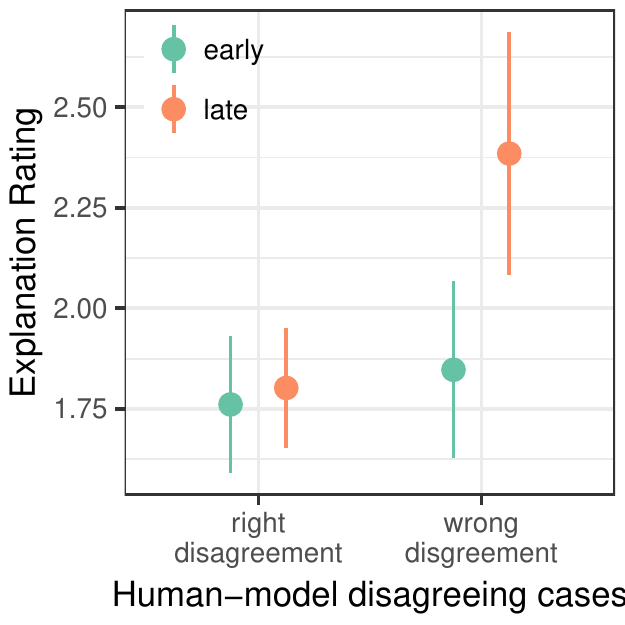}
    \caption{Explanation ratings for disagreeing instances}
    \label{fig:disagree}
\end{figure}

One may further ask whether explanation ratings provided additional information beyond the judgement expressed in the labels. For example, among cases where the participants disagreed (agreed) with the model predictions, some of them could be correct (incorrect) predictions, as compared to the ground truth. If explanation ratings could distinguish right and wrong disagreement (agreement), they could serve as additional signals that supplement instance labels. Indeed, as shown in Figure~\ref{fig:disagree}, we found that among the \textit{disagreeing instances}, participants' average explanation rating given to \textit{wrong disagreement} (the model was making the correct prediction and should not have been rejected) was higher than those to the \textit{right disagreement} ($F(1, 11)=3.12$, $p=0.10$), especially in the late stage (interactive effect between \textit{Stage} and \textit{Disagreement Correctness} $F(1, 11)=4.04$, $p=0.07$). We did not find this differentiating effect of explanation for agreeing instances. 

The above results are interesting as Teso and Kersting proposed to leverage feedback of ``weak acceptance'' to train AL ("right decision for the wrong reason"~\cite{teso2018should}), in which people agree with the system's prediction but found the explanation to be problematic. Empirically, we found that the tendency for people to give weak acceptance may be less than weak rejection. Future work could explore utilizing weak rejection to improve model learning, for example, with AL algorithms that can consider probabilistic annotations~\cite{song2018active}. 

\subsubsection{Open form feedback} 
\label{feedback}
We conducted content analysis on participants' open form answers to provide feedback, especially by comparing the ones in the CL and XAL conditions. In the XAL condition, participants had two fields as shown in Figure~\ref{fig:interface_2} to provide their feedback for the model decision and explanation. We combined them for the content analysis as some participants filled everything in one text field. In the CL condition, only the first field on model decision was shown. Two authors performed iterative coding on the types of feedback until a set of codes emerged and were agreed upon. In total, we gathered 258 entries of feedback on explanations in the XAL conditions (out of 480 trials). 44.96\% of them did not provide enough information to be considered valid feedback (e.g. simply expressing agreement or disagreement with the model). 

The most evident pattern contrasting the CL and XAL conditions is a shift from commenting on the top features to determine an income prediction to more diverse types of comments based on the explanation. For example, in the CL condition, the majority of comments were concerned with the job category to determine one's income level, such as ``\textit{Craft repair likely doesn't pay more than 80000}.'' However, for the model, job category is not necessarily the most important feature for individual decisions, suggesting that people's direct feature-level input may not be ideal for the learning model to consume. In contrast, feedback based on model explanations is not only more diverse in their types, but also covers a wider range of features. Below we discuss the types of feedback, ranked by the occurring frequency.


\begin{itemize}

    \item \textit{Tuning weights} ($N=81$): The majority of feedback focused on the weights bars in the explanation visualization, expressing disagreement and adjustment one wanted to make. E.g.,"\textit{marital status should be weighted somewhat less}". It is noteworthy that while participants commented on between one to four features, the median number of features was only one. Unlike in the CL condition where participants overly focused on the feature of job category, participants in the XAL condition often caught features that did not align with their expectation, e.g. ``\textit{Too much weight put into being married}'', or ``\textit{Age should be more negatively ranked}''. Some participants kept commenting on a feature in consecutive queries to keep tuning its weights, showing that they had a desired range in mind. 
    
    \item \textit{Removing, changing direction of, or adding features} ($N=28$): Some comments suggested, qualitatively, to remove, change the impact direction of, or add certain features. This kind of feedback often expressed surprise, especially on sensitive features such as race and gender, e.g."\textit{not sure why females would be rated negatively}", or "\textit{how is divorce a positive thing}". Only one participant mentioned \textit{adding} a feature not shown, e.g., "\textit{should take age into account}". These patterns echoed observations from prior work that local explanation heightens people's attention towards unexpected, especially sensitive features~\cite{dodge2019explaining}. We note that ``removing a feature to be irrelevant" is the feedback Teso and Kersting's AL algorithm incorporates~\cite{teso2018should}.

    \item \textit{Ranking or comparing multiple feature weights} ($N=12$) : A small number of comments explicitly addressed the ranking or comparison of multiple features, such as "\textit{occupation should be ranked more positively than marital status}".

    \item \textit{Reasoning about combination and relations of features} ($N=10$): Consistent with observation in Stumpf et al.'s study~\cite{stumpf2007toward}, some comments suggested the model to consider combined or relational effect of features--e.g., "\textit{years of education over a certain age is negligible}", or ``\textit{hours per week not so important in farming, fishing}''. This kind of feedback is rarely considered by current AL or iML systems.

    \item \textit{Logic to make decisions based on feature importance} ($N=6$): The feature importance based explanation associates the model's prediction with the combined weights of all features. Some comments ($N=6$) expressed confusion, e.g. "\textit{literally all of the information points to earning more than 80,000}" (while the base chance was negative). Such comments highlight the need for a more comprehensible design of explanations, and also indicate people's natural tendency to provide feedback on the model's overall logic.
    
    \item \textit{Changes of explanation} ($N=5$): Interacting with an online AL algorithm, some participants paid attention to the changes of explanations. For example, one participant in the condition seeing the late-stage model first noticed the declining quality of the system's rationale. Another participant commented that the weights in the model explanation ``\textit{jumps back and fourth, for the same job}''. Change of explanation is a unique property of the AL setting. Future work could explore interfaces that explicitly present changes or progress in the explanation and utilize the feedback. 
 
\end{itemize}{}

To summarize, we identified opportunities to use local explanations to elicit knowledge input beyond instance labels. By simply soliciting a rating for the explanation, additional signals for the instance could be obtained for the learning model. Through qualitative analysis of the open-form feedback, we identified several categories of input that people naturally wanted to give by reacting to the local explanations. Future work could explore algorithms and systems that utilize annotators' input based on local explanations for the model's features, weights, feature ranks, relations, and changes during the learning process. 

\section{Discussions and Future Directions}
 Our work is motivated by the vision of creating natural experiences to teach learning models by seeing and providing feedback for the model's explanations of selected instances. While the results show promises and illuminate key considerations of user preferences, it is only a starting point. To realize the vision, supporting the needs of machine teachers and fully harnessing their feedback for model explanations, requires both algorithmic advancement and refining the ways to explain and interact. Below we provide recommendations for future work of XAL as informed by the study.

\subsection{Explanations for machine teaching}
Common goals of AI explanations, as reflected in much of the XAI literature, are to support a complete and sound understanding of the model~\cite{kulesza2015principles,carvalho2019machine}, and to foster trust in the AI~\cite{poursabzi2018manipulating,cheng2019explaining}. These goals may have to be revised in the context of machine teaching. First, explanations should aim to \textit{calibrate} trust, and in general the perception of model capability, by accurately and efficiently communicating the model's current limitations. 

Second, while prior work often expects explanations to enhance adherence or persuasiveness~\cite{poursabzi2018manipulating}, we highlight the opposite problem in machine teaching, as an ``anchoring'' effect to a naive model's judgment could be counter-productive and impair the quality of human feedback. Future work should seek alternative designs to mitigate the anchoring effect. For example, it would be interesting to use a partial explanation that does not reveal the model's judgment (e.g., only a subset of top features~\cite{lai2019human}), or have people first make their own judgment before seeing the explanation.  

Third, the premise of XAL is to make the teaching task accessible by focusing on individual instances and eliciting incremental feedback. It may be unnecessary to target a complete understanding of the model, especially as the model is constantly being updated. Since people have to review and judge many instances in a row, \textit{low cognitive workload} without sacrificing the quality of feedback should be a primary design goal of explanations for XAL. One potential solution is \textit{progressive disclosure} by starting from simplified explanations and progressively provide more details~\cite{springer2019progressive}. Since the early-stage model is likely to have obvious flaws, using simpler explanations could suffice and demand less cognitive resource. Another approach is to design explanations that are sensitive to the targeted feedback, for example by only presenting features that the model is uncertain about or people are likely to critique, assuming some notion of uncertainty or likelihood information could be inferred.

While we used a local feature importance visualization to explain the model, we could speculate on the effect of alternative designs based on the results. We chose a visualization design to show the importance values of multiple features at a glance. While it is possible to describe the feature importance with texts as in ~\cite{dodge2019explaining}, it is likely to be even more cognitively demanding to read and comprehend. We do not recommend further limiting the number of features presented, since people are more inclined to critique features they see rather than recalling ones not presented. Other design choices for local explanations include presenting similar examples with the same known outcome~\cite{bien2011prototype,gurumoorthy2017protodash}, and rules that the model believes to guarantee the prediction~\cite{ribeiro2018anchors} (e.g., ``someone with an executive job above the age of 40 is highly likely to earn more than 80K``). We suspect that the example based explanation might not present much new information for feedback. The rule-based explanation, on the other hand, could be an interesting design for future work to explore, as annotators may be able to approve or disapprove the rules, or judge between multiple candidate rules~\cite{hanafi2017seer}. This kind of feedback could be leveraged by the learning model. Lastly, we fixed on local explanations for the model to self-address the \textit{why} question (intelligibility type). We believe it fits naturally with the workflow of AL querying selected instances. A potential drawback is that it requires annotators to carefully reason with the explanation for every new queried instance. It would be interesting to explore using a global explanation so that annotators would only need to attend to changes of overall logic as the model learns. But it is unknown whether a global explanation is as easy for non-AI-experts to make sense and provide feedback on. 

 There are also opportunities to develop new explanation techniques by leveraging the temporal nature of AL. One is to \textit{explain model progress}, for example by explicitly showing changes in the model logic compared to prior versions. This could potentially help the annotators better assess the model progress and identify remaining flaws. Second is to utilize \textit{explanation and feedback history} to both improve explanation presentation (e.g., avoiding repetitive explanations) and infer user preferences (e.g., how many features is ideal to present).

Lastly, our study highlights the needs to tailor explanations based on the characteristics of the teacher. People from whom the model seeks feedback may not be experienced with ML algorithms, and not necessarily possess the complete domain knowledge or contextual information. Depending on their cognitive style or the context to teach, they may have limited cognitive resources to deliberate on the explanations. These individual characteristics may impact their preferences for the level of details, visual presentation, and whether explanation should be presented at all.

\subsection{Learning from explanation based feedback}
Our experiment intends to be an elicitation study to gather the types of feedback people naturally want to provide for model explanations. An immediate next step for future work is to develop new AL algorithms that could incorporate the types of feedback presented in Section~\ref{feedback}. Prior work, as reviewed in Section~\ref{literature}, proposed methods to incorporate feedback on top features or boosting the importance of features~\cite{raghavan2006active,druck2009active,settles2011closing,stumpf2007toward}, and removing features~\cite{teso2018should,kulesza2015principles}. However most of them are for text classifiers. Since feature-based feedback for text data is usually binary (a keyword should be considered a feature or not), prior work often did not consider the more quantitative feedback shown in our study, such as tuning the weights of features, comparatively ranking features, or reasoning about the logic or relations of multiple features. While much technical work is to be done, it is beyond the scope of this paper. Here we highlight a few key observations from people's natural tendency to provide feedback for explanations, which should be reflected in the assumptions that future algorithmic work makes.

First, people's natural feedback for explanations is \textit{incremental} and \textit{incomplete}. It tends to focus on a small number of features that are most evidently unaligned with one's expectation, instead of the full set of features. Second, people's natural feedback is  \textit{imprecise}. For example, feature weights were suggested to be qualitatively increased, decreased, added, removed, or changing direction. It may be challenging for a lay person to accurately specify a quantitative correction for a model explanation, but a tight feedback loop should allow one to quickly view how an imprecise correction impacts the model and make follow-up adjustment. Lastly, people's feedback is \textit{heterogeneous}. Across individuals there are vast differences on the types of feedback, the number of features to critique, and the tendency to focus on specific features, such as whether a demographic feature should be considered fair to use~\cite{dodge2019explaining}. 

Taken together, compared to providing instance labels, feedback for model explanations can be noisy and frail. Incorporating the feedback ``as it is'' to update the learned features may not be desirable. For example, some have warned against ``local decision pitfalls''~\cite{wu2019local} of human feedback in iML that overly focuses on modifying a subset of model features, commonly resulting in an overfitted model that fails to generalize. Moreover, not all ML models are feasible to update the learned features directly. While prior iML work often builds on directly modifiable models such as regression or naïve Bayes classifiers, our approach is motivated by the possibility to utilize popular \textit{post-hoc} techniques to generate local explanations~\cite{ribeiro2016should,lundberg2017unified} for any kind of ML models, even those not directly interpretable such as neural networks. It means that an explanation could give information about how the model weighs different features but it is not directly connected to its inner working. How to incorporate human feedback for post-hoc explanations to update the original model remains an open challenge. It may be interesting to explore approaches that take human feedback as weighted signals, constraints, a part of a co-training model or ensemble~\cite{stumpf2009interacting} , or impacting the data~\cite{teso2018should} or the sampling strategy. 

A coupled aspect to make human feedback more robust and consumable for a learning algorithm is to design interfaces that scaffold the elicitation of high-quality, targeted type of feedback. This is indeed the focus of the bulk of iML literature. For example, allowing people to drag-and-drop to change the ranks of features, or providing sliders to change the feature weights, may encourage people to provide more precise and complete feedback. It would also be interesting to leverage the explanation and feedback history to extract more reliable signals from multiple entries of feedback, or purposely prompt people for confirmation of prior feedback. Given the heterogeneous nature of people's feedback, future work could also explore methods to elicit and cross-check input from multiple people to obtain more robust teaching signals.

\subsection{Explanation- and explainee-aware sampling}
Sampling strategy is the most important component of an AL algorithm to determine its learning efficiency. But existing AL work often ignores the impact of sampling strategy on annotators' experience. For example, our study showed that uncertainty sampling (selecting instance the model is most uncertain about to query) led to an increasing challenge for annotators to provide correct labels as the model matures. 

For XAL algorithms to efficiently gather feedback and support a good teaching experience, sampling strategy should move beyond the current focus on decision uncertainty to considering the explanation for the next instance and what feedback to gain from that explanation. For the machine teacher, desired properties of explanations may include easiness to judge, non-repetitiveness, tailored to their preferences and tendency to provide feedback, etc.~\cite{sokol2020explainability}. For the learning model, it may gain value from explaining and soliciting feedback for features that it is uncertain about, have not been examined by people, or have high impact on the model performance. Future work should explore sampling strategies that optimize for these criteria of explanations and explainees.

\section{Limitations}
We acknowledge several limitations of the study. First, the participants were recruited on Mechanical Turk and not held accountable for consequences of the model, so their behaviors may not generalize to all SMEs. However, we attempted to improve the ecological validity by carefully designing the domain knowledge training task and reward mechanism (participants received bonus if among 10\% performer). Second, this is a relatively small-scale lab study. While the quantitative results showed significance with a small sample size, results from the qualitative data, specifically the types of feedback may not be considered an exhaustive list. Third, the dataset has a small number of features and the model is relatively simple. For more complex models, the current design of explanation with feature importance visualization could be more challenging to judge and provide meaningful feedback for. 

\section{Conclusions}
 While active learning has gained popularity for its learning efficiency, it has not been widely considered as an HCI problem despite its interactive nature. We propose explainable active learning (XAL), by utilizing a popular local explanation method as the interface for an AL algorithm. Instead of opaquely requesting labels for selected instances, the model presents its own prediction and explanation for its prediction, then requests feedback from the human. We posit that this new paradigm not only addresses annotators' needs for model transparency, but also opens up opportunities for new learning algorithms that learn from human feedback for the model explanations. Broadly, XAL allows training ML models to more closely resemble a ``teaching'' experience, and places explanations as a central element of machine teaching. We conducted an experiment to both test the feasibility of XAL and serve as an elicitation study to identify the types of feedback people naturally want to provide. The experiment demonstrated that explanations could help people monitor the model learning progress and calibrate their trust in the teaching outcome. But our results cautioned against the adverse effect of explanations in anchoring people's judgment to the naive model's, if the annotator lacks adequate knowledge to detect the model's faulty reasoning, and the additional workload that could avert people with low Need for Cognition. Besides providing a systematic understanding of user interaction with AL algorithms, our results have three broad implications for using model explanations as the interface for machine teaching. First, we highlight the design goals of explanations applied to the context of teaching a learning model, as distinct from common goals in XAI literature, including calibrating trust, mitigating anchoring effect and minimizing cognitive workload. Second, we identify important individual factors that mediate people's preferences and reception to model explanations, including task knowledge, AI experience and Need for Cognition. Lastly, we enumerate on the types of feedback people naturally want to provide for model explanations, to inspire future algorithmic work to incorporate such feedback. 
 
\section*{Acknowledgments}
We wish to thank all participants and reviewers for their helpful feedback. This work was done as an internship project at IBM Research AI, and partially supported by NSF grants IIS 1527200 and IIS 1941613.

\bibliographystyle{ACM-Reference-Format}
\bibliography{sample-base}

\end{document}